\documentclass[aps,pre,twocolumn,floatfix,superscriptaddress,showpacs,showkeys]{revtex4-2}
\usepackage{epsf,amsmath,amssymb,verbatim,color,multirow,pifont,mathrsfs,soul,amsthm, wasysym}
\usepackage{graphicx, upgreek}
\usepackage[us,12hr]{datetime}

\begin{document}

\title{Link rewiring with local information--induced hybrid percolation transitions}
\author{Young Sul Cho}
\email{yscho@jbnu.ac.kr}
\affiliation{Department of Physics, Jeonbuk National University, Jeonju 54896, Republic of Korea}
\affiliation{Research Institute of Physics and Chemistry, Jeonbuk National University, Jeonju 54896, Republic of Korea}

\date{\today}

\begin{abstract}
When a link is occupied to restrict the growth of large clusters using the size information of a finite number of finite clusters, so-called local information, an abrupt but continuous transition is exhibited. We report here that a hybrid transition can occur if each node rewires its links to restrict the growth of large clusters using local information continuously up to a finite number of rewirings. 
For example, on a branch of a Bethe lattice with coordination number $4$, each node rewires its outgoing links to its descendants several times in ascending order of cluster size to reach a steady state. Then a hybrid transition with nontrivial critical exponents occurs as a function of the link fraction at the steady state.  
We observe this phenomenon even on a Bethe lattice without hierarchy, supporting that such a phenomenon may occur on diverse tree networks with finite degrees.
\end{abstract}

\maketitle

\section{Introduction}
Explosive percolation is a phenomenon in which the order parameter, i.e., the fraction of nodes belonging to an infinite cluster, becomes finite in an abrupt manner at a certain threshold as the fraction of occupied links increases on a random graph or a given lattice~\cite{Achlioptas:2009, explosive_phenomena, souza_nphy}. This phenomenon has been theoretically studied using diverse models and various transition natures,
such as continuous transitions with abnormal critical exponents~\cite{ziff_lattice:2010,filippo_pre:2010,fss_exp,local,tricritical,hklee,choi,dacosta_prl,dacosta_pre,grassberger,riordan,smoh:2016,eppre:2023,epprl:2023, Jan_gap1,Jan_gap2}, discontinuous transitions~\cite{ziff_ncomm, restricted, cho_science, chopre:2010, largest, gaussian, resource}, and hybrid transitions~\cite{choprl:2016, park_hybrid, choi_hybrid}. Explosive percolation is also observed in various real systems including protein networks~\cite{makse}, disordered materials~\cite{chopre:2011, nanotube, fracture, crackling}, and nanocomposites~\cite{experiment}.

A large proportion of explosive percolation models, including the first model~\cite{Achlioptas:2009}, use information on the size of some clusters to occupy a link to suppress the growth of large clusters.
For example, the first model randomly selects two unoccupied links and then occupies the one
with the smaller product of connected clusters. We note that the size information of four finite clusters is generally used to restrict the growth of large clusters until an infinite cluster is formed.
Such size information of a finite number of finite clusters is called local information, and 
information that does not fulfill this condition, such as the sizes of an infinite number of finite clusters or global cluster size distributions, is
called global information.

It is known that a continuous transition occurs when a link is occupied to suppress the growth of large clusters using local information, while a discontinuous or hybrid transition may occur if global information is used~\cite{explosive_phenomena, riordan}. Here, a hybrid transition refers to the coexistence of continuous and discontinuous transitions at a single critical point~\cite{kcore2, kcore3, baxter}.
However, this conclusion was obtained in a restricted situation where each node is not allowed to rewire its occupied links using local information, 
even though the surrounding cluster sizes generally change.
It would be natural for each node, though, to rewire its occupied links according to changing local environments, similar to how each node removes some of its occupied links as networks evolve in $k$-core percolation~\cite{kcore1} and percolation in interdependent networks~\cite{buldyrev:2010}.

In this paper, we investigate whether the transition nature changes if each node on a given lattice repeatedly rewires its occupied links to its neighbors in ascending order of cluster size.
We specifically consider the Bethe lattice, an infinite tree consistently composed of nodes with $z$ neighbors~\cite{stauffer, kim_percolation, Bethe_bond, saberi_bethe, chae2012, yscho2024}.  
First, we rigorously show that a hybrid transition occurs on a Bethe lattice branch of only outward links with $z-1=3$ by relating the rewiring process to the inverse branching process. Then we observe a hybrid transition even on a Bethe lattice of both inward and outward links with $z=4$ via simulation in Supplemental Material.

\begin{figure}[t!]
\includegraphics[width=0.8\linewidth]{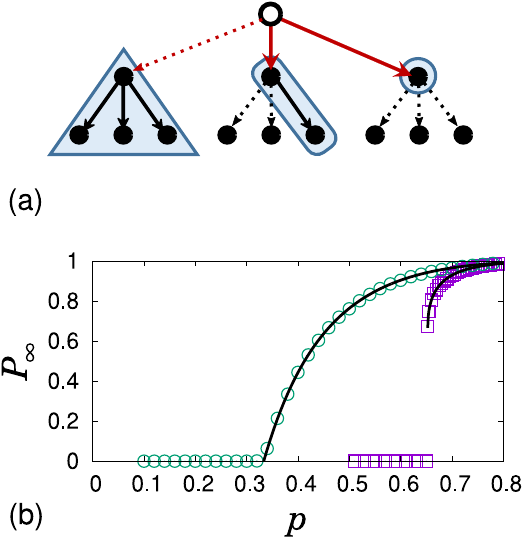}
\caption{(a) Schematic diagram of the inverse branching process following the local minimal cluster rule (LMCR) on a Bethe lattice branch with $z-1=3$ and $L=3$.
The out-degree of the root ($\Circle$) is assigned $n=2$, and the root deterministically occupies the two links attached to the smallest clusters among its neighbors on the second layer.
(b) Simulation results of $P_{\infty}$ for the inverse branching process with $(\square)$ and without $(\circ)$ the LMCR.
The results fit well with the theoretical lines.}
\label{Fig:MinBranchDet_Pinf}
\end{figure}

To computationally implement the Bethe lattice branch, we use a Cayley tree branch of $L$ layers. Each layer, indexed by $\ell=0,...,L-1$, is composed of $(z-1)^{\ell}$ nodes, and each node of the $\ell$-th layer attach outgoing links to its $z-1$ neighbors of the $(\ell+1)$-th layer.
Then the total number of nodes $N$ is given by $N=[(z-1)^L-1]/(z-2)$.
Some links are occupied following a given rule, and the cluster of a given node is defined as the set of nodes that can be reached via occupied links from the node, including the node itself.
We note that each node on the outermost layer $(\ell=L-1)$ has no links and consistently belongs to an isolated cluster of size $1$.

The order parameter $P_{\infty}$ is the probability that the root ($\ell=0$) belongs to an infinite cluster in the thermodynamic limit $N \rightarrow \infty$.
We note that the result of bond percolation $P_{\infty} = 1-(1-pP_{\infty})^{z-1}$ is obtained when each link is occupied randomly according to a given link occupation probability $p$~\cite{stauffer, kim_percolation, Bethe_bond}.

\section{Inverse branching process with the local minimal cluster rule} 
In a Cayley tree branch with given $z-1$ and $L$, we apply the local minimal cluster rule (LMCR) to an inverse branching process as follows. 
Initially, all links are unoccupied, and thus every node belongs to an isolated cluster of size $1$.  
Beginning with the second outermost layer ($\ell=L-2$), each node of the $\ell$-th layer is updated following (i,ii) as below.
\begin{itemize}
\item{(i)} The out-degree of each node is assigned among $n=0,...,z-1$ according to the probability $Q(n)=\binom{z-1}{n}p^n(1-p)^{z-1-n}$ for a given $0\leq p \leq 1$.
\item{(ii)} In the LMCR, each node occupies its outgoing links, as many as the out-degree, in ascending order of attached cluster size, as illustrated in Fig.~\ref{Fig:MinBranchDet_Pinf}(a). Links attached to clusters of the same size are occupied in random order.
\end{itemize}
This process is repeated by $\ell \rightarrow \ell-1$ until $\ell$ becomes 0.
We note that if the LMCR is not applied and each node occupies a randomly chosen $n$ number of its links in step (ii), 
the process becomes bond percolation on the Cayley tree branch.

When $z-1=3$, $P_{\infty}$ satisfies
\begin{equation}
\begin{array}{r@{}l}
P_{\infty}=Q(1)P_{\infty}^3+Q(2)\big[3P_{\infty}^2(1-P_{\infty})+P^3_{\infty}\big] \\\\
+Q(3)\big[1-(1-P_{\infty})^3\big],
\end{array}
\label{Eq:Pinf_minbranch}
\end{equation}
where each term on the r.h.s. beginning with $Q(n)$ means that at least $z-n$ outgoing neighbors of each node must be connected to infinite clusters in order for the node to be connected to an infinite cluster along $n$ occupied links. We note that this formulation is valid because the probability that each node belongs to an infinite cluster is independent among nodes on the same layer.

Then $P_{\infty}$ becomes finite discontinuously at $p_c$ as
\begin{equation}
P_{\infty} = \left\{\begin{array}{lll}
0 & \textrm{~for~} & p < p_c \\\\
P_{\infty}(p_c) + C(p-p_c)^{\beta} & \textrm{~for~} & p \rightarrow p_c^+,
\end{array}\right.
\label{Eq:Pinf}
\end{equation}
where $\beta=1/2$, $p_c \approx 0.652834$, and $P_{\infty}(p_c) \approx 0.666364$ for a $p$-independent constant $C$.
The closed form of $P_{\infty}$ obtained by solving Eq.~(\ref{Eq:Pinf_minbranch}) fits well to the simulation result, as shown in Fig.~\ref{Fig:MinBranchDet_Pinf}(b).
For the closed form of $P_{\infty}$ and the simulation method, see Sec. 1 of Supplemental Material.

For other values of $z$, $P_{\infty}$ exhibits a continuous transition as $P_{\infty} \propto (p-1/\sqrt{2})$ for $z-1=2$. 
For $z-1 > 3$, we guess that $P_{\infty}$ may exhibit a discontinuous transition irrespective of $z$, but this is not confirmed here as it is beyond the scope of this paper.
In what follows, we only consider $z-1=3$.

\section{Link rewiring process with the local minimal cluster rule}
The inverse branching process with the LMCR is somewhat deterministic in that it occupies outgoing links from the second outermost layer to inner layers in a sequential manner. Here, we show that the same phenomenon is spontaneously reached via link rewirings of each node with the LMCR.

\begin{figure}[t!]
\includegraphics[width=1.0\linewidth]{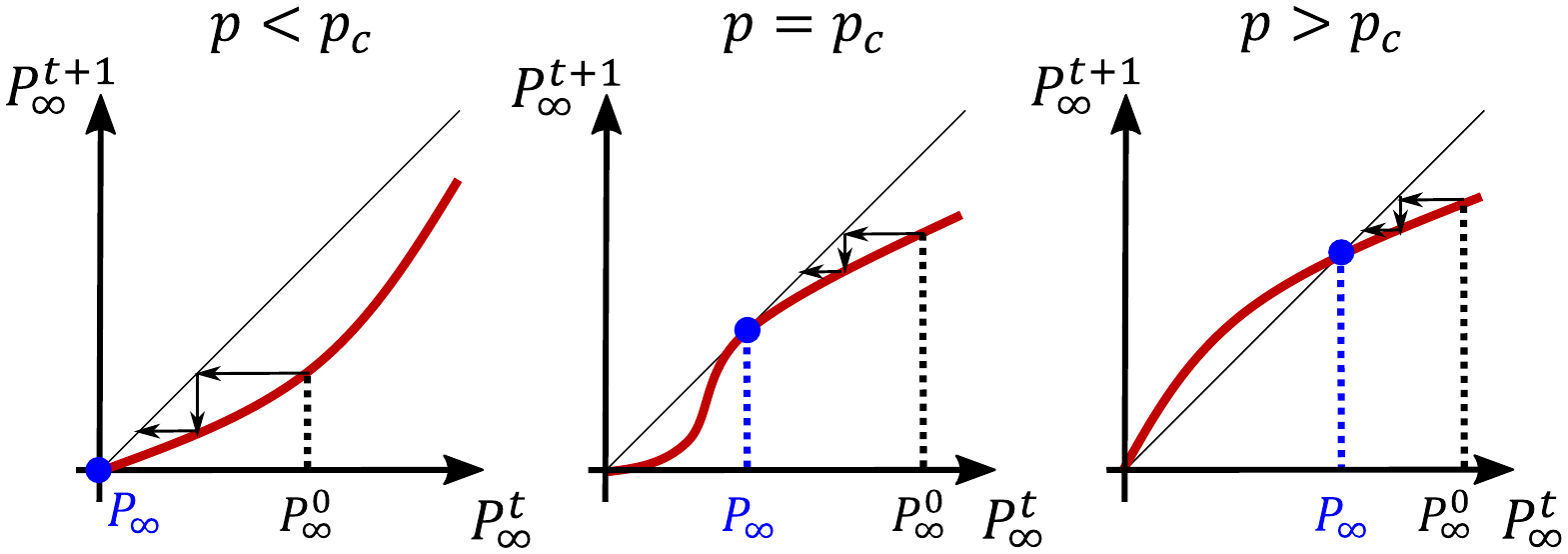}
\caption{Schematic diagram of how $P^t_{\infty}$ approaches $P_{\infty}$ of the steady state as $t$ increases for three regions of $p$. 
The thin black lines denote $P^{t+1}_{\infty}=P^t_{\infty}$ and thick red lines denote $P^{t+1}_{\infty}=f(P^t_{\infty})$.
Beginning with $P^0_{\infty}$, the order parameter shifts one step to the left following the arrows as $t$ is increased by $1$.}
\label{Fig:Schematic_saturate}
\end{figure}

\begin{figure}[t!]
\includegraphics[width=0.8\linewidth]{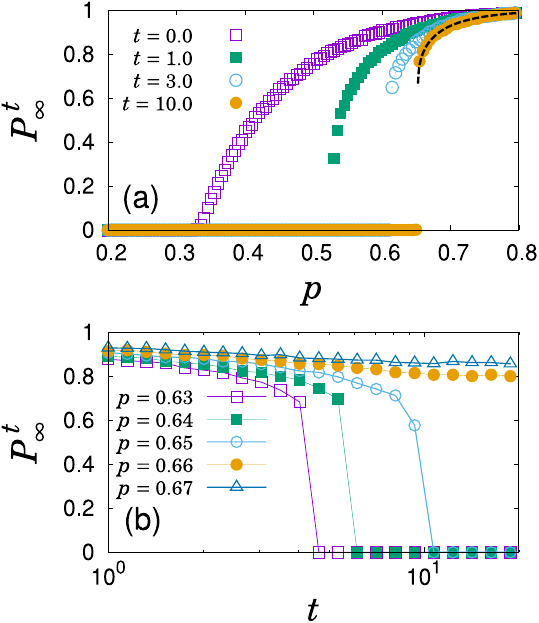}
\caption{(a) $P_{\infty}^t(p)$ vs. $p$ for several $t$. As $t$ increases, $P_{\infty}^t(p)$ 
converges to the theoretical curve of $P_{\infty}(p)$ for $p \geq p_c$ at the steady state (dashed line). (b) $P_{\infty}^t(p)$ vs. $t$ for several $p$. $P_{\infty}^t(p)$ drops to $0$ at finite $t$
for $p<p_c$, while it converges to $P_{\infty}(p)>0$ for $p>p_c$.} 
\label{Fig:MinBranchRewire_Pinf}
\end{figure}

At the beginning we occupy each link with probability $p$ on a Cayley tree branch with $z-1=3$ and given $L$.
At each time step, we randomly select one node to rewire its occupied links following the LMCR. 
It is guaranteed that the cluster size monotonically decreases during the rewiring process after the initial occupation of links.
The rewiring process continues until it reaches a steady state, where the size of every cluster no longer decreases.

With this setup we investigate via simulation the behavior of $P_{\infty}(p)$ as the number of rewirings per node increases. The radius $L$ is very small compared to $N$ on the Cayley tree branch, and therefore it is not feasible to estimate $P_{\infty}$ via direct implementation of the model using a finite $L$. 
To estimate $P_{\infty}$ via simulation, we use a modified method, described in Sec. 2.1 of Supplemental Material.
As shown in Fig.~\ref{Fig:MinBranchRewire_Pinf}(a), $P_{\infty}$ approaches Eq.~(\ref{Eq:Pinf}) after a finite number of rewirings per node.

We understand this phenomenon as follows. We denote the number of rewirings per node as $t$ and
the value of $P_{\infty}$ at $t$ as $P^{t}_{\infty}$. It is guaranteed that $P^t_{\infty}$ monotonically decreases with $t$ 
because the size of the cluster the root belongs to monotonically decreases with $t$. We assume that every node belongs to an infinite cluster with probability $P^{t}_{\infty}$ consistently at $t$. 
Then, $P^{t}_{\infty}$ updates as $P^{t+1}_{\infty}=f(P^t_{\infty})$, 
where $f(x)$ is given by $P_{\infty}=f(P_{\infty})$ in Eq.~(\ref{Eq:Pinf_minbranch}).
Beginning with the result of bond percolation $P^{0}_{\infty}$, $P_{\infty}^t$ decreases to $P_{\infty}$ as $t$ increases. A schematic diagram for the saturation process is illustrated in Fig.~\ref{Fig:Schematic_saturate} for different ranges of $p$.
In this manner, we confirm that $P_{\infty}$ of the steady state of the rewiring process is the same as that of the inverse branching process.
We note that each network generated by the inverse branching process does not require any further link rewirings following the LMCR (see also Sec. 2.2 of Supplemental Material).

According to the above explanation using Fig.~\ref{Fig:Schematic_saturate}, the number of rewirings needed for $P_{\infty}^t$ to converge to $P_{\infty}$ of the steady state
may increase as $p$ approaches $p_c$ from below, because $y=x$ is tangent to $y=f(x)$ at $x=P_{\infty}(p_c)$ when $p=p_c$.
This prediction is confirmed by the result that $P_{\infty}$ drops to $0$ at a larger number of rewirings per node as $p$ approaches $p_c$ from below, as shown in Fig.~\ref{Fig:MinBranchRewire_Pinf}(b). It will be interesting to study how the dropping point increases as $p$ approaches $p_c$ from below in the future.

We now discuss whether the discontinuity of $P_{\infty}(p_c)$ is indeed induced by local information for $p<p_c$, unlike in previous explosive percolation models.
When $1/3 < p < p_c$, $P^0_{\infty}>0$ and each node should calculate cluster sizes extensive to $N$ with a finite probability at the beginning of the rewirings,
which actually requires global information.
To avoid this, the model can be modified to occupy links randomly, where each node rewires several times following the LMCR after each link occupation. 
It is confirmed that the modified model reveals the same phenomenon---using local information---when the fraction of occupied links is treated as $p$, as shown in Sec. 2.3 of Supplemental Material.

We also check whether a continuous transition occurs if links are occupied following the LMCR without link rewiring~\cite{chae2012, yscho2024}. Specifically, each node is selected in random order with an assigned out-degree of $n=0,...,z-1$ according to probability $Q(n)$, and the node follows the LMCR without further rewiring. In this case, after the root occupies its links, there is no opportunity to rewire even if the order of neighboring cluster sizes changes due to occupied links in outer shells. A detailed explanation on the origin of the continuous transition is presented in Sec. 3 of Supplemental Material. This result confirms that the rewiring process is essential to reveal a discontinuity of $P_{\infty}(p_c)$ of Eq.~(\ref{Eq:Pinf}).

\section{Criticality after discontinuity of the order parameter} 
A hybrid transition is the coexistence of a discontinuity of $P_{\infty}$ and critical phenomena of the cluster size distribution at $p_c$. To examine whether a transition is hybrid or not, we discuss the cluster size distribution $P_s$, which is the probability that the root belongs to a cluster of finite size $s$.

For this we regard the steady states of the rewiring process as the results of the inverse branching process. Then
$P_s$ for $s \geq 1$ follows 
\begin{widetext}
\begin{equation}
\begin{array}{r@{}l}
P_s = Q(0)\delta_{s1} + Q(1)\Big[P_{s-1}\Big(1-\sum_{s'=1}^{s-2}P_{s'}\Big)^2 + 2P_{s-1}\Big(1-\sum_{s'=1}^{s-1}\Big)^2 
+ P_{s-1}^2 \Big(1-\sum_{s'=1}^{s-1}P_{s'}\Big)\Big] \\\\ + Q(2)\Big[P^3_{\frac{s-1}{2}}\delta_{\frac{s-1}{2} \lfloor \frac{s-1}{2}\rfloor} + 3\sum_{i=1}^{\lfloor\frac{s-1}{2}\rfloor}P_iP^2_{s-1-i} 
+ 3\sum_{i+j=s-1}P_iP_j\Big(1-\sum_{k=\textrm{max}(i,j)}P_k\Big)\Big]+Q(3)\sum_{i+j+k=s-1}P_iP_jP_k, 
\end{array}
\label{Eq:cdist}
\end{equation}
\end{widetext}
where each term on the r.h.s. beginning with $Q(n)$ is the probability that the sum of the cluster sizes connected to the root through $n$-occupied links is $s-1$,
and $\lfloor x \rfloor$ is the greatest integer less than or equal to $x$.
We note that Eq.~(\ref{Eq:Pinf_minbranch}) is obtained by applying the relation $P_{\infty} = 1-\sum_{s=1}^{\infty}P_s$ to Eq.~(\ref{Eq:cdist}).

\begin{figure}[t!]
\includegraphics[width=0.8\linewidth]{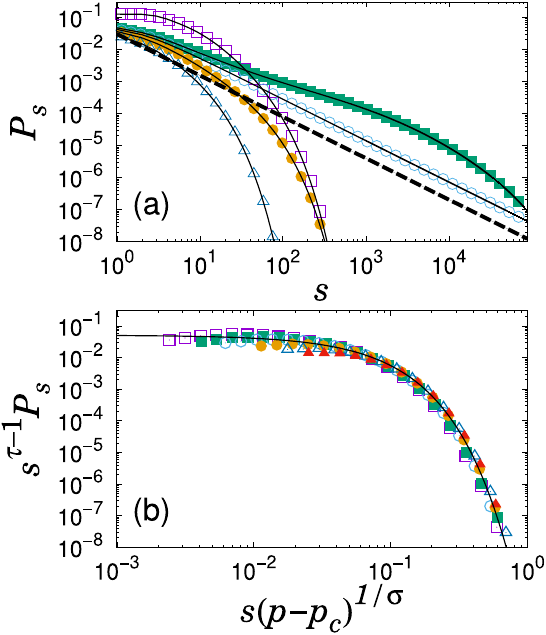}
\caption{(a) $P_s$ obtained via numerical integration of Eq.~(\ref{Eq:cdist}) (lines) and simulation (symbols) for $p=0.5, 0.64, 0.652834, 0.67, 0.7$ from top to bottom, where the slope of the dashed line is $-1.293$. To obtain the simulation data, we use $L=11$ and set each node on the outermost layer to belong to a cluster of size $s$ with probability $P_s$ as obtained through numerical integration. The data collapse of
$P_s$ is obtained via simulation for $p=0.68 (\square), 0.69 (\blacksquare), 0.7 (\Circle), 0.72 (\bullet), 0.74 (\triangle), 0.76 (\blacktriangle)$ onto the scaling function. The line is proportional to $\textrm{exp}(-22x)$.} 
\label{Fig:MinBranchRewire_cdist}
\end{figure}

We obtain $P_s$ for $1 \leq s<s^*$ satisfying $P_{s^*} = 10^{-20}$ via numerical integration of Eq.~(\ref{Eq:cdist}) for various $p$.
In the entire range of $0<p<1$, $P_s$ clearly exhibits hybrid transition behavior, as shown in Fig.~\ref{Fig:MinBranchRewire_cdist}(a).
At first, $P_s$ decreases exponentially in the range of small $p \ll p_c$. Then $P_s$ has a bump in the range of large $s \gg 1$ as $p$ approaches $p_c$ from below.
At $p_c$, the bump is eliminated by a discontinuous jump of the order parameter up to $P_{\infty}(p_c)$, and $P_s$ shows a clear power law behavior, $P_s \propto s^{-\tau}$, with the nontrivial exponent $\tau = 1.293 \pm 0.001$. The analytical value of $\tau$ may be derived using Eq.~(\ref{Eq:cdist}). As $p \rightarrow p_c^+$, $P_s$ shows the critical behavior of a continuous transition as $P_s \propto s^{-\tau}\textrm{exp}(-x)$ for $x \propto s(p-p_c)^{1/\sigma}$, satisfying $(\tau-2)/\sigma = \beta=1/2$ as shown in Fig.~\ref{Fig:MinBranchRewire_cdist}(b). We also check in Fig.~\ref{Fig:MinBranchRewire_cdist} that $P_s$ of the steady state obtained through simulation gives the same result.
These findings confirm that the percolation transition is a hybrid transition.

\section{Discussion}
We review the mechanism of the discontinuity of $P_{\infty}$ at $p_c$ discovered in this paper and explore general systems in which a phenomenon induced by the same mechanism might be observed.
On a Cayley tree branch, the cluster of the root should expand outward through each layer to increase its size.
At the steady state, every node belonging to each layer suppresses expansion through its outgoing links using local information,
such that the growth of the cluster of the root is suppressed globally to reveal discontinuity of the order parameter. 
This is guaranteed because the Cayley tree branch is, by definition, a tree, and the degree of every node is finite.

We now examine whether such a phenomenon can occur even on a Bethe lattice without hierarchy, meaning where each node attaches outgoing links to $z$ neighbors.
In this case, the size of a cluster can increase unlike the branch, but $P_{\infty}$ monotonically decreases. Therefore, the existence of a steady state in which $P_{\infty}$ no longer decreases is also guaranteed here. On the Bethe lattice, we observe that a discontinuity of $P_{\infty}(p_c)$ with the criticality of $P_s$ already emerges after several rewirings before reaching a steady state. Detailed arguments and results of the Bethe lattice are presented in Sec. 4 of Supplemental Material. 
Accordingly, we expect that such a phenomenon can be observed in diverse tree lattices of finite degree.
But in the absence of hierarchy, it is difficult to find an analytically tractable steady state, such as the inverse branching process. Additional studies are therefore needed for a rigorous analysis.

Finally, we discuss the relationship between the present result and previous explosive percolation models. 
Diverse studies have been conducted on occupying links using local information in random graphs~\cite{er} or various lattices and measuring the abnormal critical exponents of abrupt continuous transitions.
A network generated on a random graph is a tree before the critical point, and thus the problem on a random graph can also be treated as a problem on a random tree lattice up to the transition point. Therefore, if rewiring among neighbors using the LMCR is continuously applied to each node on the lattices used in previous studies, 
diverse results are expected on how phase transitions change depending on the lattice structures.

\section{Acknowledgements}
This work was supported by the National Research Foundation (NRF) of Korea, grant no. NRF-2020R1F1A1061326.
Data will be made available
on reasonable request.

\newpage

\section*{Supplemental Material}

\section*{1. Inverse branching process with the local minimal cluster rule}
\subsection*{Simulation method to implement $P_{\infty}$}
In this section, we describe the simulation method to implement $P_{\infty}$ of the inverse branching process with the local minimal cluster rule (LMCR).
At the beginning of the iteration, $P^{(1)}_{\infty}=1$ is given. In the $i$-th iteration, an extensive number of independent realizations of the following process (i--iv) are performed.
\begin{itemize}
\item[(i)] On the Cayley tree branch with $z-1=3$ and given $L$, each node on the outermost layer is determined to belong to an infinite cluster extending outwards with the probability $P^{(i)}_{\infty}$ or
to be a finite cluster with the probability $1-P^{(i)}_{\infty}$.
\item[(ii)] If a cluster contains at least one node on the outermost layer belonging to an infinite cluster, it is an infinite cluster; otherwise, it is a finite cluster. Here, the size of a finite cluster is calculated by consistently treating the size of the finite clusters that the outermost nodes belong to as 1.
\item[(iii)] Beginning with the second outermost layer ($\ell=L-2$), (a) each node of the $\ell$-th layer is assigned to occupy $n=0,...,z-1$ number of links
according to probability $Q(n)$. (b) Each node of the $\ell$-th layer then occupies its links to finite clusters in ascending order of their sizes, and remaining links are occupied to infinite clusters.
This process is repeated by $\ell \rightarrow \ell-1$ until $\ell$ becomes 0.
\end{itemize}
The fraction of independent realizations where the root belongs to an infinite cluster becomes $P^{(i+1)}_{\infty}$ for the $(i+1)$-th iteration. Iterations continue until $P^{(i)}_{\infty}$ converges.

\subsection*{Closed form of $P_{\infty}$}

The closed form of $P_{\infty}$ of Eq.~(1) in the main text is given by 

\begin{align}
P_{\infty}= \notag \\ &\frac{-9p^2+12p^3-\sqrt{12p-48p^2+40p^3+45p^4-72p^5+24p^6}}{6p-24p^2+20p^3} \notag \\
       &\textrm{for}~~p_c\leq p \leq 1
\end{align}
where $0 < p_c < 1$ satisfies $12p_c-48p_c^2+40p_c^3+45p_c^4-72p_c^5+24p_c^6=0$ and $0 \leq P_{\infty}(p_c) \leq 1$.
The numerical values of $p_c$ and $P_{\infty}(p_c)$ are $p_c \approx 0.652834$ and $P_{\infty}(p_c) \approx 0.666364$.

\section*{2. Link rewiring process with the local minimal cluster rule}
\subsection*{Simulation method to implement $P_{\infty}$ at steady states}
\label{sec:branchmethod}

In this section, we describe the simulation method to implement $P_{\infty}$ of the link rewiring process with the LMCR in the branch.  
At the beginning of the iteration, $P^{(1)}_{\infty}=1$ is given. In the $i$-th iteration, an extensive number of independent realizations of the following process (i--iv) are performed. 
\begin{itemize}
\item[(i)] Each link is occupied with a probability $p$ on the Cayley tree branch with $z-1=3$ and given $L$. 
\item[(ii)] Each node on the outermost layer is determined to belong to an infinite cluster extending outwards with the probability $P^{(i)}_{\infty}$ or
to be a finite cluster with the probability $1-P^{(i)}_{\infty}$.
These determinations are fixed during the entire process.
\item[(iii)] If a cluster contains at least one node on the outermost layer belonging to an infinite cluster, it is an infinite cluster; otherwise, it is a finite cluster. Here, the size of a finite cluster is calculated by consistently treating the size of the finite clusters that the outermost nodes belong to as 1.
\item[(iv)] At each time step, a node is selected randomly. Links of the node are rewired to finite clusters in ascending order of their sizes, and remaining links are rewired to infinite clusters.
This rewiring process continues until it reaches a steady state where the fraction of nodes belonging to infinite clusters no longer decreases.
\end{itemize}
The fraction of independent realizations where the root belongs to an infinite cluster becomes $P^{(i+1)}_{\infty}$ for the $(i+1)$-th iteration. Iterations continue until $P^{(i)}_{\infty}$ converges to $P_{\infty}$.

\subsection*{Equivalence between steady states and networks generated by the inverse branching process}

\begin{figure}[t!]
\includegraphics[width=0.8\linewidth]{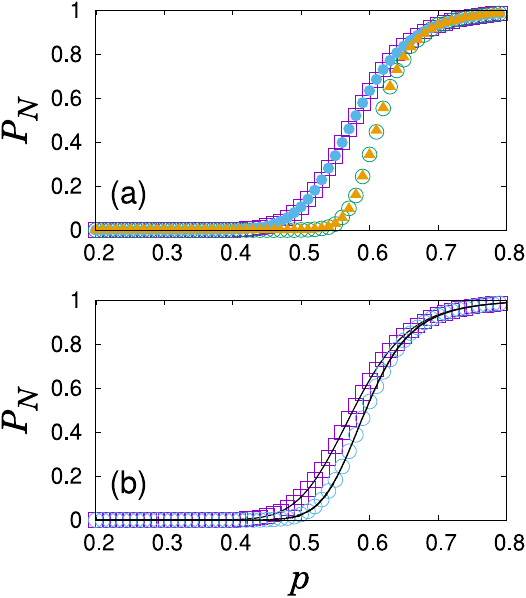}
\caption{(a) $P_N$ of the inverse branching process for $L=8 (\Square), 14(\Circle)$ and
the link rewiring process for $L=8 (\bullet), 14(\blacktriangle)$. (b) $P_N$ of the link
rewiring process for $L=8(\Square), 10(\Circle)$ and the process of link rewirings after each random occupation of a link for $L=8, 10$ (lines) from the left.} 
\label{Fig:RewireGlobalDet_Pinf}
\end{figure}

In Fig.~\ref{Fig:RewireGlobalDet_Pinf}(a), we measure $P_N$ of both the link rewiring process and the inverse branching process with the LMCR via direct implementation of the model 
on a Cayley tree branch with a finite $L$, where $P_N$ is the fraction of independent realizations where the cluster of the root includes at least one node on the outermost layer.
As shown in Fig.~\ref{Fig:RewireGlobalDet_Pinf}(a), $P_N$ of both models perfectly fit to each other irrespective of $N$, which means that both models generate the same $P_{\infty}$
in the thermodynamic limit $N \rightarrow \infty$.

\subsection*{Equivalence between steady states and networks generated by the process of link rewirings after each random occupation of a link}

To obtain $P_N$ of the process of link rewirings with the LMCR after each random occupation of a link, we perform an extensive number of independent realizations of the following process (i--iii).
\begin{itemize}
\item{(i)} At the beginning, all links are unoccupied on the Cayley tree branch with $z-1=3$ and given $L$.
\item{(ii)} At each time step, one link is occupied randomly. Then nodes with at least one occupied link rewire their links to neighbors in ascending order of cluster sizes in random order.
\item{(iii)} Step (ii) is repeated until all links on the Cayley tree branch are occupied.
\end{itemize}

We check that $P_N$ obtained in this manner fits to that of the steady states of the rewiring process irrespective of $N$, as shown in Fig.~\ref{Fig:RewireGlobalDet_Pinf}(b).

\section*{3. Continuous transition of link occupation with the local minimal cluster rule without link rewiring}
\label{sec:cont}

We first show that $P_{\infty}$ obtained using the following simulation method exhibits a continuous transition.
At the beginning of the iteration, $P^{(1)}_{\infty}=1$ is given. In the $i$-th iteration, an extensive number of independent realizations of the following process (i--iv) are performed. 
\begin{itemize}
\item[(i)] A Cayley tree branch with $z-1=3$ and given $L$ is used. 
\item[(ii)] Each node on the outermost layer is determined to belong to an infinite cluster extending outwards with the probability $P^{(i)}_{\infty}$ or
to be a finite cluster with the probability $1-P^{(i)}_{\infty}$.
These determinations are fixed during the entire process.
\item[(iii)] If a cluster contains at least one node on the outermost layer belonging to an infinite cluster, it is an infinite cluster; otherwise, it is a finite cluster. Here, the size of a finite cluster is calculated by consistently treating the size of the finite clusters that the outermost nodes belong to as 1.
\item[(iv)] Each node is selected in random order and the selected node is assigned to occupy $n=0,...,z-1$ number of its links according to probability $Q(n)$.
The node occupies its links to finite clusters in ascending order of their sizes, and remaining links are occupied to infinite clusters.
This process continues until all nodes are selected to occupy their links.
\end{itemize}
The fraction of independent realizations where the root belongs to an infinite cluster becomes $P^{(i+1)}_{\infty}$ for the $(i+1)$-th iteration. Iterations continue until $P^{(i)}_{\infty}$ converges to $P_{\infty}$.

\begin{figure}[t!]
\includegraphics[width=1.0\linewidth]{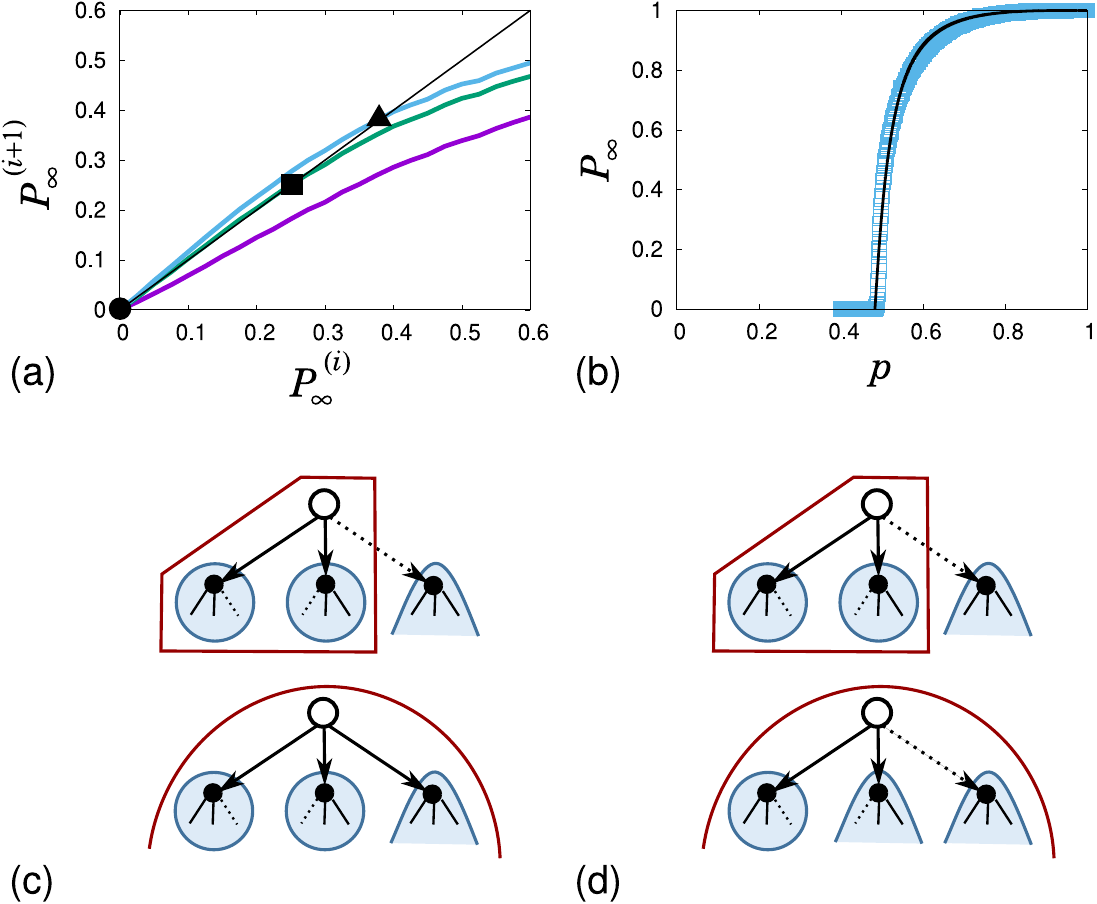}
\caption{(a) The thin solid line is $y=x$ and the thick solid lines are $P_{\infty}^{(i+1)}$ obtained via an extensive number of independent realizations following (i--iv) in Sec.~\ref{sec:cont} for a given $P_{\infty}^{(i)}$. The data are obtained using $L=10$ at $p=0.48, 0.493, 0.5$ from below. The $y$ value of a stable fixed point is $P_{\infty}$ at each $p$, where $P_{\infty}$ becomes nonzero continuously as $p$ exceeds $0.48$. The symbols mark $P_{\infty}$ at $p=0.48 (\CIRCLE), 0.493 (\blacksquare), 0.5 (\blacktriangle)$. (b) $P_{\infty}$ obtained via simulation $(\square)$ with $L=10$ and numerical integration with $C_0=0.093, C_1=0.8148$, along with $p_c=0.483$ (line). 
(c,d) Schematic diagrams of the cases corresponding to the first (c) and second (d) terms of the r.h.s. of Eq.~(\ref{eq:minbranch}). In the upper panels, two occupied (solid) arrows of the root $(\Circle)$ are attached to finite clusters, and the unoccupied (dotted) arrow of the root is attached to an infinite cluster initially. 
In the lower panel of (c), the root belongs to an infinite cluster as the dotted arrow is occupied. In the lower panel of (d), the root belongs to an infinite cluster as one of the finite clusters becomes an infinite cluster.} 
\label{Fig:MinBranch_Pinf}
\end{figure}

$P_{\infty}$ obtained using this method exhibits a continuous transition, as shown in Fig.~\ref{Fig:MinBranch_Pinf}(a) and (b). 
To understand the origin of the continuous transition, we assume that each node occupies links to finite clusters in random order regardless of their size. 
This assumption explains well the origin of the continuous transition, described in more detail below.

We treat this model as a dynamical process of occupying links, where $p$ becomes the fraction of occupied links.
Specifically, at each time step, a node is selected with a probability proportional to its number of unoccupied links, and it occupies a link attached to a finite cluster if one exists; otherwise, it occupies a link attached to an infinite cluster.
Under the assumption in the preceding paragraph, the rate equation of $P_{\infty}$ for this process is given by 
\begin{equation}
\frac{\partial P_{\infty}}{\partial p} = \frac{1}{1-p}F_1(p) + \frac{1}{1-P_{\infty}}\frac{\partial P_{\infty}}{\partial p}F_2(p),
\tag{S2}
\label{eq:minbranch}
\end{equation}
where the first term of the r.h.s. is the rate of all cases in which the cluster of the root becomes an infinite cluster as a link of the root is occupied, and 
the second term of the r.h.s. is the rate of all cases in which the cluster of the root becomes an infinite cluster as the cluster of a neighbor becomes an infinite cluster.
Representative cases corresponding to these terms are depicted in Fig.~\ref{Fig:MinBranch_Pinf}(c) and (d).
We note that $1/(1-p)$ and $1/(1-P_{\infty})\partial P_{\infty}/\partial p$ on the r.h.s. are rates for a fixed unoccupied bond being occupied and for a fixed finite cluster becoming an infinite cluster, respectively,

Considering all cases corresponding to both terms, all rate equations to be solved numerically can be derived with Eq.~(\ref{eq:minbranch}).
We solve the rate equations for $p_c \leq p$ using the best-fit values of $C_0, C_1, p_c$ with the initial conditions $C_1+2C_2+3C_3=1, C_0+C_1+C_2+C_3=1$, and $P_{\infty}(p_c)=0$,
where $C_n$ denotes the probability that the root occupies $n$ links at $p_c$ in the rate equations. $P_{\infty}$ obtained via numerical integration of Eq.~(\ref{eq:minbranch}) fits 
well with the simulation result as shown in Fig.~\ref{Fig:MinBranch_Pinf}(b). Additionally, $P_{\infty}(p_c)=0$ of the rate equation indicates a continuous transition, which explains why the model exhibits a continuous transition.

\section*{4. Link rewiring with the local minimal cluster rule on the Bethe lattice}
\subsection*{Model on a finite-sized Cayley tree and example}
We introduce the model on a Cayley tree with $z=4$ neighbors and $L$ layers to implement simulation. Each layer is indexed by $\ell=0,...,L-1$, where the layer $\ell=0$ is the root and each layer $\ell \geq 1$ is composed of $z(z-1)^{\ell-1}$ nodes. 
The root has outgoing links attached to $z$ neighbors on the $\ell=1$ layer, and nodes on the outermost layer have no links.
Each node of other layers $1 \leq \ell \leq L-2$ has one outgoing link attached to a node of the $(\ell-1)$-th layer and $z-1$ outgoing links attached to $z-1$ nodes of the $(\ell+1)$-th layer. 
Definitions of $P_{\infty}$ and cluster are the same as in the branch. 
We note that each node on the outermost layer consistently belongs to an isolated cluster of size $1$.

At the beginning, each link is occupied with probability $p$. At each time step, we select one node of the layers $0 \leq \ell \leq L-2$ randomly, and it rewires occupied links to neighbors in ascending order based on the number of reachable nodes outward along those neighbors.
This rewiring process continues up to a given total number of rewirings per node.

\begin{figure}[t!]
\includegraphics[width=1.0\linewidth]{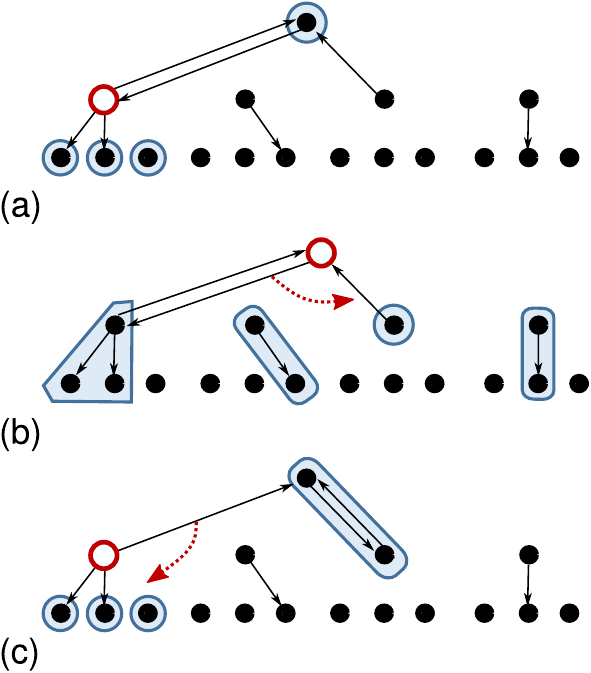}
\caption{Example of the rewiring process on a Bethe lattice with $L=3$. Unoccupied bonds are not depicted for clear description. (a) A node ($\Circle$) on the $\ell=1$ layer is randomly selected to rewire. Four neighboring subsets that are reachable outward from the node are highlighted in blue. The node maintains its occupied links because the size of all subsets is equal to $1$. (b) A root ($\Circle$) is randomly selected to rewire. Four neighboring subsets that are reachable outward from the root are highlighted in blue. The root rewires its occupied link to the smallest one following the dotted arrow. (c) A node ($\Circle$) on the $\ell=1$ layer is randomly selected to rewire. Four neighboring subsets that are reachable outward from the node are highlighted in blue. The root rewires its occupied link to the smallest one following the dotted arrow.} 
\label{Fig:MinBetheRewireSchematic}
\end{figure}

Fig.~\ref{Fig:MinBetheRewireSchematic} depicts an example of the rewiring process. This example shows that the rewiring of a node affects the cluster sizes of its neighbors on the outer layer, unlike the branch, which proves the absence of hierarchy in the Bethe lattice.

\subsection*{Existence of a steady state}
\label{sec:steadystate}

The example in Fig.~\ref{Fig:MinBetheRewireSchematic} also shows that the size of a cluster can increase during the rewiring process, unlike the branch. Specifically,
in (a), the empty circle belongs to a cluster of size $4$, but after
the root rewires in (b), the empty circle belongs to a cluster of size $5$. For example, a cluster can increase when one node in a pair of nodes with directed links to each other rewires its link to another node outside of the pair, 
as in (b). In this case, the opposite node of the pair generally has a larger cluster as the link heading toward it turns outward.
On the contrary, we note that there is no pair connecting each other, and thus the cluster size monotonically decreases in the branch.

According to this discussion, the size of a cluster can increase in the Bethe lattice by rewiring with the LMCR. Therefore, unlike in the branch, a steady state in which the sizes of all clusters no longer decrease is generally not observed. However, a finite cluster does not become an infinite cluster via link rewiring, because a link may not be rewired to an infinite cluster from a finite cluster. Therefore, $P_{\infty}$ monotonically decreases during the rewiring process, and a steady state in which the fraction of nodes belonging to an infinite cluster no longer decreases should exist even in the Bethe lattice.
We checked that $P_{\infty}$ monotonically decreases following the simulation method in Sec.~\ref{sec:Bethemethod} but failed to observe a steady state, because of the computation time.

\subsection*{Simulation method to implement $P_{\infty}$ after a finite number of rewirings per node}
\label{sec:Bethemethod}

In this section, we describe the simulation method to implement $P_{\infty}$ of the link rewiring process with the LMCR in the Bethe lattice. 
Here we introduce $A$, which is the probability that the cluster of the root extends outward infinitely through one fixed neighbor.
At the beginning of the iteration, $A^{(1)}=1$ is given. In the $i$-th iteration, an extensive number of independent realizations of the following process (i--iv) are performed. 
\begin{itemize}
\item[(i)] Each link is occupied with a probability $p$ on a Cayley tree with $z=4$ and given $L$. 
\item[(ii)] Each node on the outermost layer is determined to belong to an infinite cluster extending outward with probability $A^{(i)}$ or
to be an isolated cluster of size $1$ with probability $1-A^{(i)}$. If a cluster contains at least one node on the outermost layer belonging to
an infinite cluster, it is an infinite cluster; otherwise, it is a finite cluster.
\item[(iii)] At each time step, a node is selected randomly and it calculates the subsets of nodes that are reachable outward along its neighbors. If a subset contains at least one node on the outermost layer belonging to an infinite cluster, it is an infinite subset; otherwise, it is a finite subset.
\item[(iv)] The node occupies its links to finite subsets in ascending order of their size, and the remaining links are occupied to infinite subsets.
This process continues up to a given number of total rewirings per node.
\end{itemize}
The fraction of independent realizations where the subsets reachable outward along one fixed neighbor from the root is infinite becomes $A^{(i+1)}$ for the $(i+1)$-th iteration. Iterations continue until $A^{(i)}$ converges to $A$ and the fraction of independent realizations where the root belongs to an infinite cluster becomes $P_{\infty}$.

\subsection*{Results}

\begin{figure*}[t!]
\includegraphics[width=1.0\linewidth]{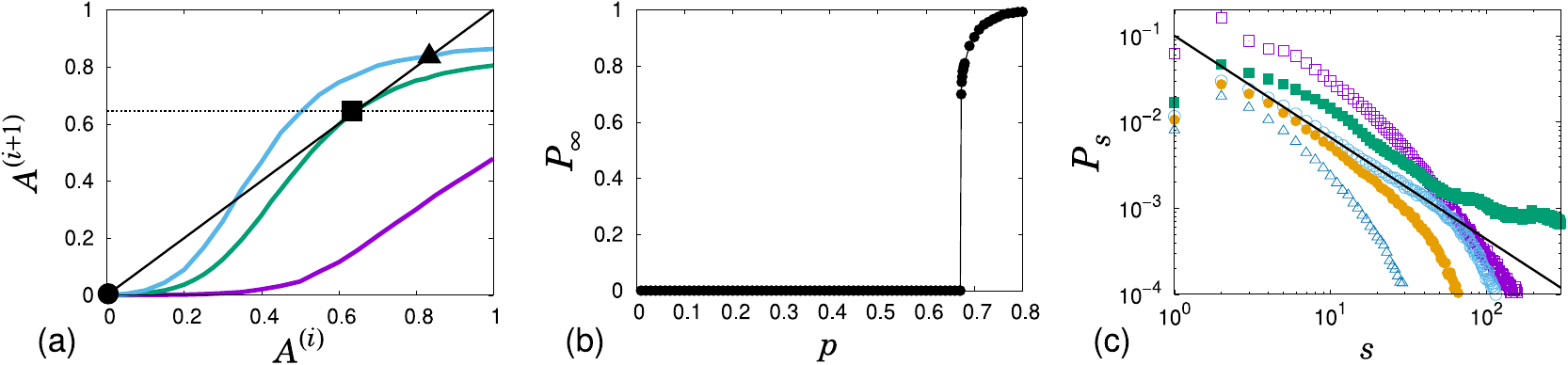}
\caption{(a) The thin black line is $y=x$ and the thick colored lines are $A^{(i+1)}$ obtained via an extensive number of independent realizations following (i--iv) in Sec.~\ref{sec:Bethemethod} for a given $A^{(i)}$.
The data are obtained by performing up to $10$ rewirings per node using $L=8$ at $p=0.62, 0.673, 0.69$ from below. The $y$ values of the symbols are $A$ at $p=0.62 (\CIRCLE), 0.673 (\blacksquare), 0.69 (\blacktriangle)$, where the dotted line indicates $0.65$. (b) $P_{\infty} (\bullet)$ obtained using the method in Sec.~\ref{sec:Bethemethod} with $L=8$. (c) $P_s$ obtained via simulation for $p=0.5, 0.64, 0.674, 0.68, 0.7$ from top to bottom, where slope of the line is $-1.18$.} 
\label{Fig:BethefAPinf}
\end{figure*}

In Fig.~\ref{Fig:BethefAPinf}(a), $A^{(i+1)}$ is tangential to $y=x$ near $A^{(i+1)} = 0.65$ at $p=0.673$, which means that $A(p)$ jumps from $0$ to $A(p_c) \approx 0.65$ discontinuously at $p_c \approx 0.673$,
when $10$ rewirings per node have been performed. We checked that the estimated $A(p_c)$ does not depend on $L$ for different values of $L$. As a result, we observe a discontinuity of $P_{\infty}$ near $p_c$, as shown in Fig.~\ref{Fig:BethefAPinf}(b). 
In Fig.~\ref{Fig:BethefAPinf}(c), we observe the criticality of $P_s \propto s^{-\tau}$ near $p_c$, and this result supports that the transition here is a hybrid transition. However, we use the approximation that finite cluster of each node on the outermost layer is an isolated cluster of size $1$. To analyze the critical behaviors of $P_s$ in more detail, specific properties of $P_s$ need to be obtained, which is beyond the current scope.

\end{document}